\documentclass[a4paper,11pt]{article}
\usepackage{pos}
\usepackage[compat=1.1.0]{tikz-feynman}

\title{Indications for New Higgs Bosons 
}

\author*[a,b]{Andreas Crivellin}
\author[c,d]{Saiyad Ashanujjaman}
\author[e]{Sumit Banik}  
\author[f,g]{Siddharth P.~Maharathy}
\author[h]{Guglielmo Coloretti}

\affiliation[a]{Universitat Autònoma de Barcelona, 08193 Bellaterra, Barcelona}

\affiliation[b]{ICREA, Instituci\'o Catalana de Recerca i Estudis Avan\c{c}ats,\\
Passeig de Llu\'{\i}s Companys 23, 
08010 Barcelona, Spain}

\affiliation[c]{Institut f\"ur Theoretische Teilchenphysik, Karlsruhe Institute of Technology, Engesserstra\ss e 7, D-76128 Karlsruhe, Germany}
\affiliation[d]{Institut f\"ur Astroteilchenphysik, Karlsruhe Institute of Technology, Hermann-von-Helmholtz-Platz 1, D-76344 Eggenstein-Leopoldshafen, Germany}

\affiliation[e]{SLAC National Accelerator Laboratory, Stanford University,\\ Stanford, California 94039, USA}

\affiliation[f]{School of Physics and Institute for Collider Particle Physics, University of the Witwatersrand, Johannesburg, Wits 2050, South Africa}
\affiliation[g]{Indian Institute of Science Education and Research Pune, Dr.~Homi Bhabha Road, Pune 411008, India}

\affiliation[h]{INFN, Sezione di Bologna,\\ Via Irnerio 46, 40126 Bologna, Italy}

\emailAdd{andreas.crivellin@cern.ch}
\emailAdd{saiyad.ashanujjaman@kit.edu}
\emailAdd{banik@stanford.edu}
\emailAdd{siddharth.prasad.maharathy@cern.ch}
\emailAdd{guglielmo.coloretti@bo.infn.it}

\abstract{After the Higgs discovery, the question of whether particles beyond those of the Standard Model exist is more pressing than ever. In this context, the scalar sector is particularly promising, since it lies at the core of the internal problems of the Standard Model, while extensions of it allow us to resolve them and can provide explanations for Dark matter, non-zero neutrino masses, inflation etc. In these proceedings, we review the indications for new Higgs bosons at the electroweak scale with masses of $\approx$95\,GeV and $\approx$152\,GeV. These excesses are most significant in the di-photon channel but are supported by weaker-than-expected limits in other decay modes. While for the 95\,GeV candidate the production mechanism is mostly unknown, the (hypothetical) 152\,GeV Higgs is dominantly produced in association with leptons, $(b)$ jets and missing energy, pointing towards the Drell-Yan production of an $SU(2)_L$ triplet with $Y=0$. Interestingly, this model predicts $t\to H^\pm b$ with $H^\pm\to WZ$, which resembles the signature of $t\bar{t}Z$ production in the Standard Model and is in fact preferred by current data. Finally, we investigate the possibility that the significant tensions between the Standard Model predictions and the measurements in differential top-quark distributions are due to contamination from new physics involving both the 152\,GeV and the 95\,GeV scalar.}

\FullConference{Proceedings of the Corfu Summer Institute 2025 "School and Workshops on Elementary Particle Physics and Gravity" (CORFU2025)\\
27 April - 28 September, 2025\\
Corfu, Greece\\}

\begin{document}
\maketitle

\section{Introduction}

The Standard Model (SM) of particle physics is the currently accepted theory describing the known fundamental constituents of matter and their interactions. It has been extensively tested and verified within the last decades by a plethora of measurements~\cite{ParticleDataGroup:2022pth} and the discovery of the Higgs boson at the Large Hadron Collider (LHC) in 2012 at CERN~\cite{Aad:2012tfa,Chatrchyan:2012ufa} confirmed the existence of its last missing ingredient; the only known fundamental scalar particle.

However, the SM cannot account for the non-vanishing neutrino masses observed via oscillations (in its original and minimal form without right-handed neutrinos) nor for the astrophysical evidence for Dark Matter, the accelerated expansion of the universe (Dark Energy) or the indications for an inflationary era. Furthermore, the Sakharov criteria~\cite{Sakharov:1967dj} for generating a matter-antimatter asymmetry are not fulfilled in the SM. Therefore, it should be extended by new particles and new interactions.

In addition to these experimental evidence for new physics, there are theoretical shortcomings of the SM: The Higgs interactions with quarks and leptons contain as many as 13 free parameters (9 fermion masses, 3 mixing angles and one complex phase). The problem of explaining the hierarchical structure with small mixing angles and only the top quark having an order-one coupling to the Higgs while the electron one is of the order of $10^{-6}$, is known as the {flavour puzzle}. It is also not understood why the Higgs mass and the related scale of electroweak (EW) symmetry breaking are many orders of magnitude smaller than the Planck mass (the hierarchy problem), and it is not clear if the vacuum of the SM scalar potential is absolutely stable~\cite{Degrassi:2012ry}. Note that all of these issues are related to the Higgs sector.

\begin{figure}[t!]
	\centering
\includegraphics[width=1\textwidth]{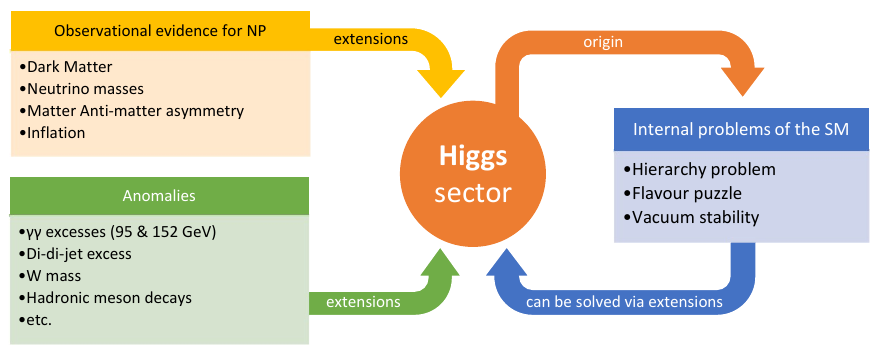}
	\caption{The Higgs sector is not only at the heart of the problems of the SM; extensions of it can also solve these issues, and experimental evidence for Dark Matter and Neutrino masses points towards new scalars. Furthermore, there are anomalies, i.e.~statistically significant indications for new Higgs bosons (particularly in the di-photon channel at 95\,GeV and 152\,GeV.}\label{figLFUV}
\end{figure}

However, the Higgs sector is not only at the heart of the problems of the SM, but extensions of it can solve them. For example, an extended scalar sector can lead to a strong first-order phase transition and provide new sources of CP violation (as required for EW scale Baryogenesis~\cite{Kuzmin:1985mm}), a scalar triplet with $Y=1$ can generate neutrino masses~\cite{Schechter:1980gr} and a neutral scalar could be DM. Furthermore, new scalars (such as supersymmetric top partners) can solve the hierarchy problem and a slow-rolling scalar field can lead to inflation.

Finally, anomalies~\cite{Crivellin:2023zui}, i.e., indications for new physics in the form of deviations from the SM predictions, point towards the existence of new scalar bosons. Therefore, extensions of the Higgs sector are particularly well motivated (see Fig.~\ref{figLFUV}) and a discovery might be possible with run-3 data. In these proceedings, we will focus on the di-photon excesses at 95\,GeV and 152\,GeV, discuss possible explanations in terms of concrete models as well as possible connections between them.

\section{Indications for New Higgses at the Electroweak Scale}


On the experimental side, as long as the participation of the new Higgses in EW symmetry breaking is small, the bounds from precision observables, Higgs signal strength~\cite{Langford:2021osp, ATLAS:2021vrm} and LHC searches are relatively weak. With few exceptions (like the type-II 2HDM) most models can be realised at the EW scale without being in conflict with experimental bounds due to the high background levels at the LHC. The constraints are particularly weak for {high-multiplicity final states} for which the invariant mass either cannot be fully reconstructed due to {missing energy}, or where it is difficult due to the limited jet-energy resolution of the ATLAS and CMS detectors in combination with the large QCD background. This is e.g.~the case for scalar bosons with EW scale masses decaying to $W$ bosons or top quarks, in particular, if they are intermediate off-shell states.

\begin{figure*}[t!]
\centering
\includegraphics[scale=0.6]{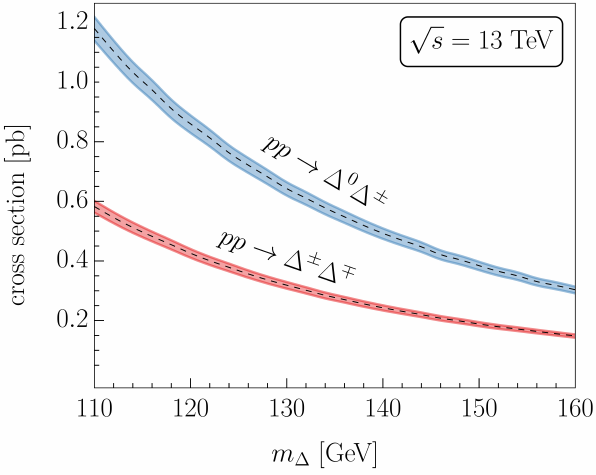}
\hspace{0.3cm}
\includegraphics[scale=0.6]{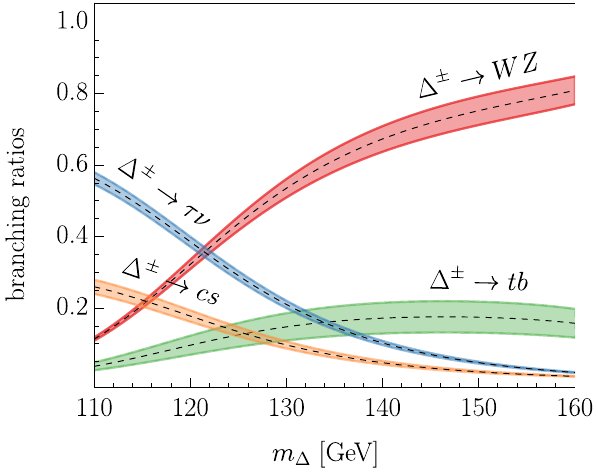}
\caption{Left: Production cross-section for $pp\to \Delta^0 \Delta^\pm$ and $pp\to \Delta^\pm \Delta^\mp$ as a function of the triplet mass. Right: Dominant branching ratios of the charged component $\Delta^\pm$ as a function of its mass.}
\label{fig:xsec}
\end{figure*}

\subsection{The 95\,GeV Candidate}

LEP reported an excess with a local significance of 2.3$\sigma$ at $\approx$98\,GeV in Higgsstrahlung, i.e.~$e^+e^-\to ZS^\prime$ with $S^\prime\to bb$~\cite{LEPWorkingGroupforHiggsbosonsearches:2003ing}. In this mass region, CMS found an excess of $2.9\sigma$ (locally) in the di-photon channel at 95\,GeV~\cite{CMS:2023yay}. At a similar mass, ATLAS observed a smaller but consistent excess of 1.7$\sigma$~\cite{ATLAS:2023jzc}, also in the di-photon channel. These indications for a new scalar are supported by a di-tau excess reported by CMS~\cite{CMS:2022goy}\footnote{Even though no dedicated analysis of the relevant mass region in the di-tau channel from ATLAS exists, no excess is indicated in the corresponding SM Higgs analysis~\cite{ATLAS:2022yrq}.} and weaker-than-expected limits in the $WW$ channel~\cite{CMS:2022uhn,ATLAS:2022ooq,Coloretti:2023wng}. Combining these results, one obtains a global significance of $3.4\sigma$~\cite{Bhattacharya:2023lmu} and signal strength w.r.t.~to a hypothetical SM Higgs of the same mass of
\begin{equation}
 \begin{aligned}\mu_{\gamma \gamma}^{\exp } &=0.27\pm 0.1\,, \qquad \mu_{b b}^{\exp } &=0.12 \pm 0.06\,, \qquad \mu_{\tau \tau}^{\exp } &=0.3 \pm 0.3\,, \qquad \mu_{W W}^{\exp } &=15 \pm 7\,.\end{aligned}   
\end{equation}

\subsection{The 152\,GeV Candidate}

First indications of a scalar with a mass around 150\,GeV were contained within the so-called ``multi-lepton anomalies", i.e.~LHC signatures with leptons, missing energy and ($b$-)jets (see Refs.~\cite{Fischer:2021sqw} for reviews), pointing towards its associated production ~\cite{vonBuddenbrock:2016rmr,vonBuddenbrock:2017gvy,Buddenbrock:2019tua,vonBuddenbrock:2020ter}. Later, Ref.~\cite{Crivellin:2021ubm} showed that the side-bands of the SM Higgs boson analyses of ATLAS~\cite{ATLAS:2020pvn,Aad:2020ivc,Aad:2021qks} and CMS~\cite{Sirunyan:2021ybb,Sirunyan:2020sum,CMS:2018nlv,Sirunyan:2018tbk} suggest the presence of a narrow scalar resonance with a mass of $\approx$151\,GeV, subsequently confirmed by Refs.~\cite{ATLAS:2023omk,ATLAS:2024lhu} such that in a simplified model analysis, the $5\sigma$ threshold has been crossed~\cite{Bhattacharya:2025rfr}.

\section{The $\Delta$SM}

Going beyond the simplified model approach for the 152\,GeV candidate, let us look in more detail at the results of Refs.~\cite{ATLAS:2023omk,ATLAS:2024lhu} which performed extensive analyses of the associated production of the SM Higgs in the di-photon channel. From the presented side-bands 
one can test the hypothesis of the presence of an additional scalar. Among the multiple associated production channels considered ($\gamma\gamma+X$), noticeable excesses are seen for $X=\tau,\ell,{\rm MET}$ and $\ell b$, while for $X=2\tau,2\ell$ and $t_{\rm lep}$, no surplus of events is observed. 

This pattern suggests the Drell-Yan production~\cite{FileviezPerez:2008bj} of the neutral 152\,GeV scalar together with the charged component of the corresponding multiplet~\cite{Banik:2024ftv}. In particular, the decay $H^\pm\to \tau \nu$ results in a single $\tau$ lepton, rather than a di-$\tau$ final state. It also produces a single charged lepton through the leptonic decay of the $\tau$, but not a dilepton signature, and is accompanied by moderate missing energy arising from the associated neutrinos. Similarly, $H^\pm\to t^*b$ leads to two $b$ jets and a lepton, thus contributing to the $\ell b$ channel but not to the $t_{\rm lep}$ one (disregarding acceptance and efficiency effects). 

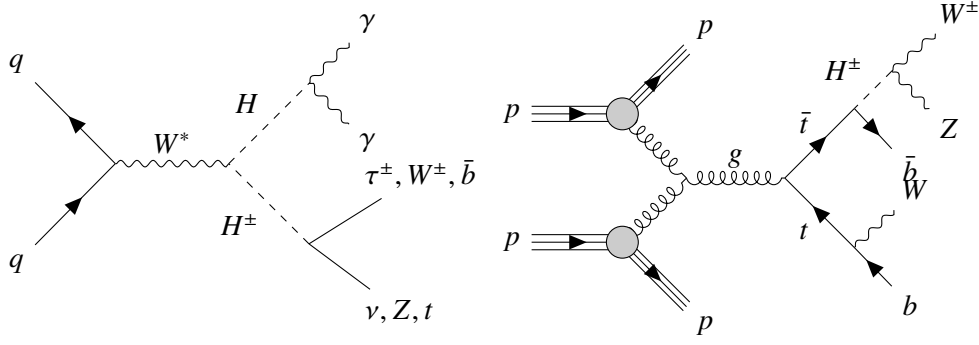
\begin{figure}[t]
    \centering
    ~~~
    \begin{tikzpicture}[baseline=(current bounding box.center)]
        \begin{feynman}
            \vertex (a);
            \vertex [above left=1.5cm of a] (c) {$q$};
            \vertex [below left=1.5cm of a] (d) {$q$};
            \vertex [right=1.5cm of a] (b) ;
            \vertex [above right=1.5cm of b] (e);
            \vertex [below right=1.5cm of b] (f);    
            \vertex [above right=0.75cm of e] (i) {$\gamma$};
            \vertex [below right=0.75cm of e] (j) {$\gamma$};
            \vertex [above right=0.85cm of f] (k) {$\tau^\pm, W^\pm, \bar b$};
            \vertex [below right=0.85cm of f] (l) {$\nu, Z, t$};
            \diagram{
                (d) -- [fermion] (a) -- [fermion] (c);
                (a) -- [boson, edge label=$W^*$] (b);
                (f) -- [scalar, edge label=$H^\pm$] (b) -- [scalar, edge label=$H$] (e);
                (j) -- [boson] (e) -- [boson] (i);
                (l) --  (f) --  (k);     };
        \end{feynman}
    \end{tikzpicture}
    \begin{tikzpicture}[baseline=(current bounding box.center)]
\begin{feynman}
\vertex (a);
\vertex [above left=1.2cm of a] (c);
\vertex [above=0.1cm of c] (cu); 
\vertex [below=0.1cm of c] (cd);

\vertex [below left=1.2cm of a] (d); 
\vertex [above=0.1cm of d] (du); 
\vertex [below=0.1cm of d] (dd);
\vertex [left=1.2cm of c] (p1) {$p$}; 
\vertex [left=1.2cm of cu] (p1u); 
\vertex [left=1.2cm of cd] (p1d);

\vertex [left=1.2cm of d] (p2) {$p$}; 
\vertex [left=1.2cm of du] (p2u); 
\vertex [left=1.2cm of dd] (p2d);

\vertex [above right=1.2cm of c] (pp1) {$p$}; 
\vertex [above right=1.15cm of cu] (pp1u); 
\vertex [above right=1.26cm of cd] (pp1d); 

\vertex [below right=1.2cm of d] (pp2) {$p$}; 
\vertex [below right=1.26cm of du] (pp2u); 
\vertex [below right=1.14cm of dd] (pp2d); 

\vertex [right=1.3cm of a] (b) ;

\vertex [above right=1.3cm of b] (e);
\vertex [below right=1.3cm of b] (f);

\vertex [above right=0.7cm of e] (j); 
\vertex [above right=0.7cm of j] (m){$W^\pm$};
\vertex [below right=0.7cm of j] (n){$Z$};
\vertex [below right=0.7cm of e] (i) {$\bar b$};

\vertex [above right=0.7cm of f] (k) {$W$};
\vertex [below right=0.7cm of f] (l) {$b$};

\diagram{
(p1) -- [fermion] (c) -- [gluon] (a), (p1u) -- (cu), (p1d) --  (cd),
(p2) -- [fermion] (d) -- [gluon] (a),  (p2u) -- (du), (p2d) -- (dd),

(c) -- [fermion] (pp1), (cu) -- (pp1u), (cd) -- (pp1d),
(d) -- [fermion] (pp2), (du) -- (pp2u), (dd) -- (pp2d),

(a) -- [gluon, edge label=$g$] (b),

(f) -- [fermion, edge label=$t$] 
(b) -- [fermion, edge label=$\bar t$] (e),

(j) -- [scalar, edge label'=$H^\pm$] (e) -- [fermion] (i),
(l) -- [fermion] (f) -- [boson] (k),

(j) -- [boson] (m),
(j) -- [boson] (n)
};
\end{feynman}
\node[draw, circle, minimum size=12pt, inner sep=0pt, fill=gray!40] at (c) {};
\node[draw, circle, minimum size=12pt, inner sep=0pt, fill=gray!40] at (d) {};
\end{tikzpicture}
    \caption{Left: Feynman diagram showing the Drell-Yan production process $pp\to W^*\to H^\pm H$ with $H\to\gamma\gamma$ and $H^\pm\to tb$, $H^\pm\to \tau\nu$ and $H^\pm\to WZ$. Right: Feynman diagram for $pp \to t\bar{t}$ with $t \to H^\pm b$ and $H^\pm \to W^\pm Z$, leading to a $t\bar{t}Z$-like signature.}
    \label{fig:Feynman1}
\end{figure}

\begin{figure}[t!]
\centering
\includegraphics[scale=0.525]{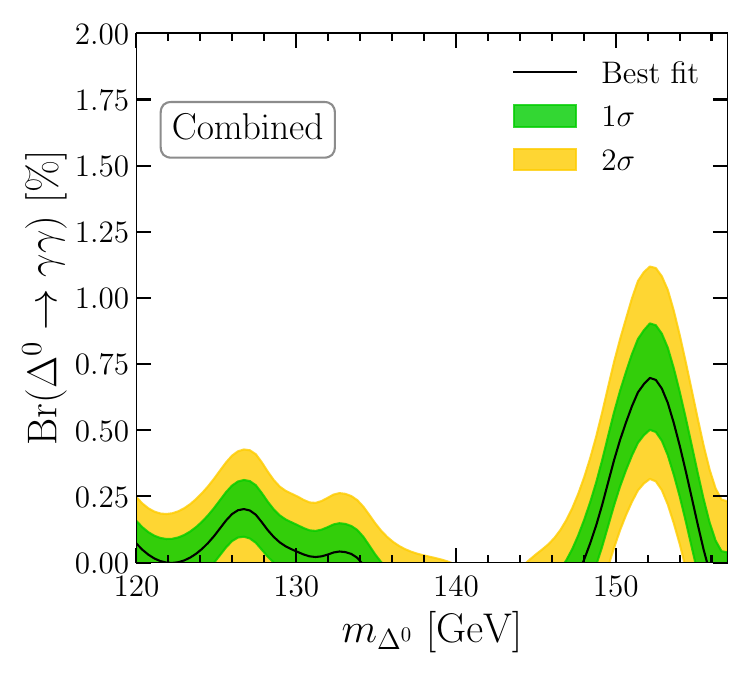}
\includegraphics[width=0.47\textwidth]{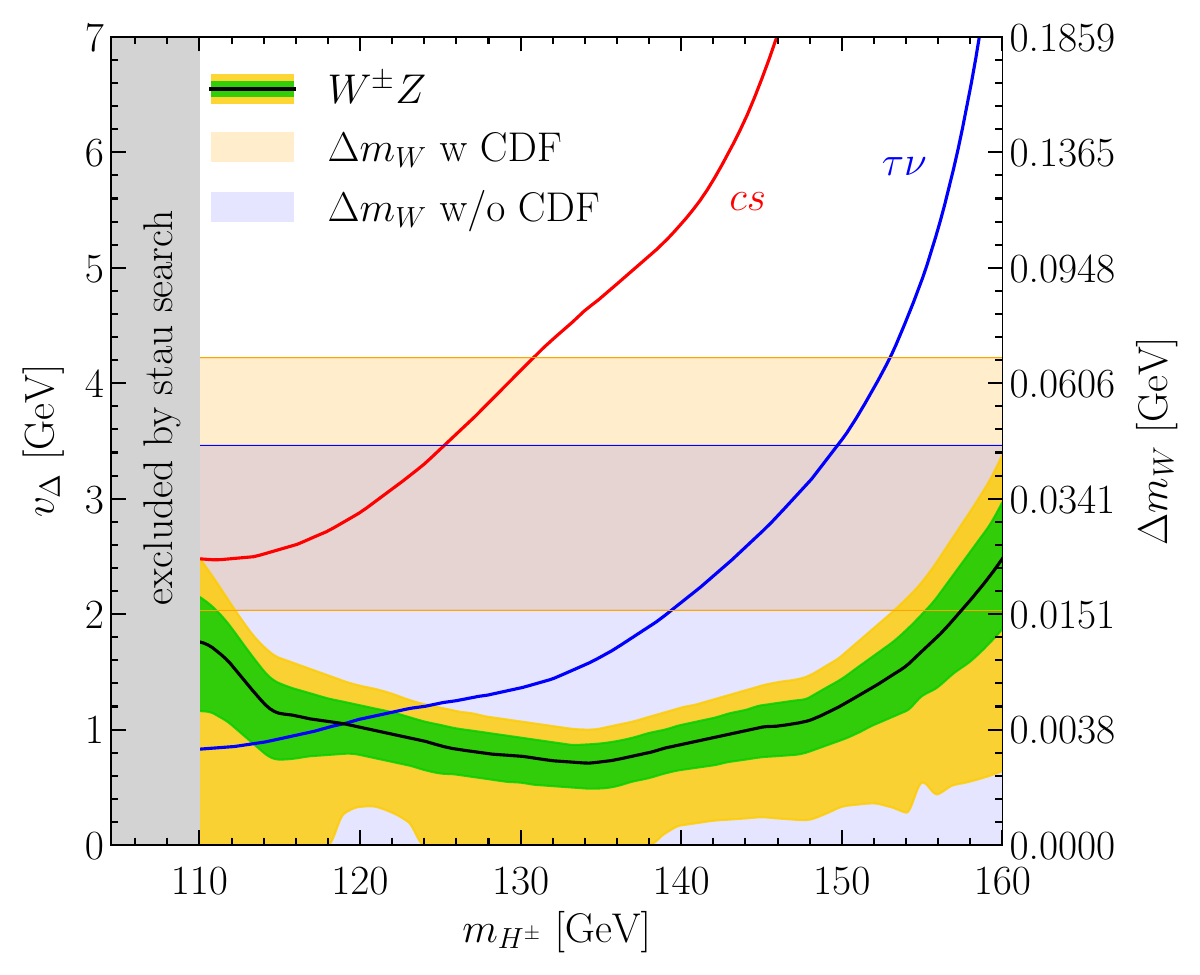}
\caption{{Left: Statistical combination of the 8 relevant channels including their correlations within the $\Delta$SM resulting in a significance of $\approx 4\sigma$ at $m_\Delta\approx152$\,GeV.} Right: Preferred range for $v_\Delta$ from $t\bar tZ$ measurement interpreted within the $\Delta$SM model as a function of $m_{H^\pm}$. The gray band is excluded by stau searches and the area above the blue (red) line by LHC searches for $t\to H^\pm$ with $H^\pm\to \tau\nu (cs)$ at 95\% CL. The light orange (blue) region shows the preferred shift in the $W$-mass from the global electroweak fit, including (excluding) the CDF-II measurement at $2\sigma$.}
\label{fig:exclusion_comb}
\end{figure}

For the Drell-Yan production, the new Higgs field, whose neutral component is the 152\,GeV candidate, must transform non-trivially under $SU(2)_L$. Because the LHC limits on multiply charged Higgses are quite stringent~\cite{Butterworth:2022dkt}, this leaves only an $SU(2)_L$ doublet with hypercharge $Y=1/2$ and the $SU(2)_L$ triplet with $Y=0$. As the latter does not decay to $ZZ$ but only to $WW$ (in the absence of mixing with the SM Higgs), which is important for the explanation of the $t\bar t$ differential distributions, we will consider here this case.

The SM supplemented by an $SU(2)_L$ triplet with $Y=0$ is called the $\Delta$SM. It is a very simple and predictive extension of the SM, containing (in addition to the SM particle content) a neutral Higgs $\Delta^0$ and a charged Higgs $\Delta^\pm$. The mass splitting between the two Higgses is proportional to their vacuum expectation value (VEV). As this VEV $v_\Delta$ only contributes (positively) to the $W$ mass but not to the $Z$ mass, the model violates custodial symmetry and $v_\Delta$ can be at most of the order of a GeV in order to respect EW precision data, rendering the charged and neutral Higgs quasi-degenerate in mass. Their production cross section at the LHC is thus determined by EW gauge invariance and a single mass $m_\Delta$. Furthermore, the dominant decay widths of $\Delta^\pm$ are all proportional to $v_\Delta^2$ thus, the branching ratios are also only a function of $m_\Delta$. This is illustrated in Fig.~\ref{fig:xsec}.

Therefore, we can perform a global fit to Br$(\Delta^0\to\gamma\gamma)$ as a function of $m_\Delta$, taking into account all signal regions of Refs.~\cite{ATLAS:2023omk,ATLAS:2024lhu}. The result is shown in Fig.~\ref{fig:exclusion_comb} (left), displaying a $\approx 4\sigma$ preference for a non-zero NP signal at 152\,GeV.

\subsection{Connection to $t\bar t Z$ distributions}

Keeping in mind that the charged and the neutral component of the Higgs triplet are quasi-degenerate in mass, the decay $t\to \Delta^\pm b$ is kinematically allowed if $m_{\Delta^0}=152$\,GeV. As the dominant decay mode of $\Delta^\pm$ in this mass region is $WZ$, the process $pp\to t\bar t\to Wb \Delta^\pm \bar b$ generates a $t\bar{t}Z$-like signal (see Fig.~\ref{fig:Feynman1}, right). This enables us to use the measurements of differential $t\bar tZ$ and $tWZ$ cross sections by CMS~\cite{CMS:2024mke} and ATLAS~\cite{ATLAS:2023eld} to search for this process~\cite{Ashanujjaman:2025una}.

In Fig.~\ref{fig:exclusion_comb}, we show the limit on $v_\Delta$ obtained from our recast~\cite{Ashanujjaman:2025una} and compare it to those from dedicated LHC searches for $t\to H^\pm b$ in the $cs$~\cite{ATLAS:2024oqu} (red) and $\tau\nu$~\cite{ATLAS:2024hya} (blue) modes, together with the constraints from the world $W$-mass fit with and without the CDF-II measurement~\cite{LHC-TeVMWWorkingGroup:2023zkn, ParticleDataGroup:2024cfk} (light orange and light blue). The recasted limit from the stau searches~\cite{CMS:2022syk} (grey) excludes charged Higgs masses below 110\,GeV~\cite{Ashanujjaman:2024lnr}. We see that our limits from the $WZ$ channel are stronger than those from the dedicated $cs$ and $\tau\nu$ searches and surpass the electroweak precision constraints across the entire mass range. Furthermore, note the $\approx2\sigma$ preference for a NP signal around 150\,GeV, which is in agreement with the predictions of the $\Delta$SM, given that $\Delta^0$ explains the di-photon excess at 152\,GeV.

\section{Differential $t\bar t$ distributions}

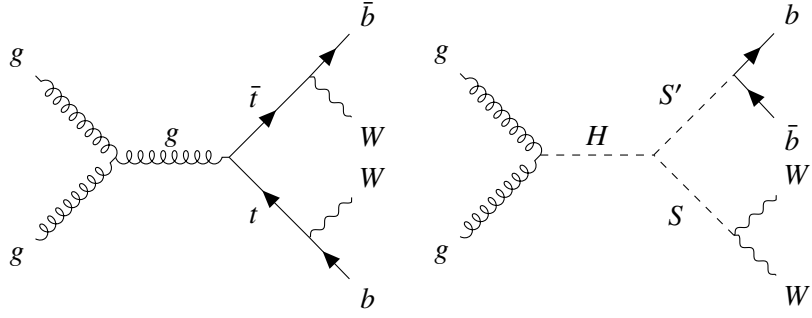
\begin{figure}[t!]
    \centering
    \begin{tikzpicture}[baseline=(current bounding box.center)]
            \begin{feynman}
            \vertex (a);
            \vertex [above left=1.5cm of a] (c) {$g$};
            \vertex [below left=1.5cm of a] (d) {$g$};
            \vertex [right=1.5cm of a] (b) ;
            \vertex [above right=1.5cm of b] (e);
            \vertex [below right=1.5cm of b] (f);
            
            \vertex [above right=0.75cm of e] (i) {$\bar b$};
            \vertex [below right=0.75cm of e] (j) {$W$};
            \vertex [above right=0.75cm of f] (k) {$W$};
            \vertex [below right=0.75cm of f] (l) {$b$};
            \diagram{
                (d) -- [gluon] (a) -- [gluon] (c);
                (a) -- [gluon, edge label=$g$] (b);
                (f) -- [fermion, edge label=$t$] 
                (b) -- [fermion, edge label=$\bar t$] (e);
                (j) -- [boson] (e) -- [fermion] (i);
                (l) -- [fermion] (f) -- [boson] (k);
            };
        \end{feynman}
            \end{tikzpicture}\quad
                \begin{tikzpicture}[baseline=(current bounding box.center)]
        \begin{feynman}
            \vertex (a);
            \vertex [above left=1.5cm of a] (c) {$g$};
            \vertex [below left=1.5cm of a] (d) {$g$};
            \vertex [right=1.5cm of a] (b) ;
            \vertex [above right=1.5cm of b] (e);
            \vertex [below right=1.5cm of b] (f);
            
            \vertex [above right=0.75cm of e] (i) {$b$};
            \vertex [below right=0.75cm of e] (j) {$\bar{b}$};
            \vertex [above right=0.75cm of f] (k) {$W$};
            \vertex [below right=0.75cm of f] (l) {$W$};
            \diagram{
                (d) -- [gluon] (a) -- [gluon] (c);
                (a) -- [scalar, edge label=$H$] (b);
                (f) -- [scalar, edge label=$S$] 
                (b) -- [scalar, edge label=$S^\prime$] (e);
                (j) -- [fermion] (e) -- [fermion] (i);
                (l) -- [boson] (f) -- [boson] (k);
            };
        \end{feynman}
    \end{tikzpicture}
    \caption{Feynman diagrams showing the leading SM contribution to top pair production and decay as well as the NP signal in our benchmark model, contaminating the measurement of $t\bar t$ differential distributions.}
    \label{fig:feynmantt}
\end{figure}

Let us now be more ambitious and try to find a connection between the 152\,GeV and the 95\,GeV Higgs candidates. This is possible by considering the process $pp\to H\to (S(152)\to WW)(S^\prime(95)\to b\bar b)$ (see Fig.~\ref{fig:feynmantt}) which has the same final state as $t\bar t$ production and decay with in the SM. 

The cleanest observables related to this process are the differential lepton distributions (assuming leptonic $W$ decays), where significant tensions with the SM predictions have been observed. They are most pronounced in the invariant mass of the lepton pair ($m_{e\mu}$) and the angle between the leptons ($\Delta \phi_{e\mu}$), where a significant surplus of events at low masses and low angles exists~\cite{ATLAS:2023gsl}.\footnote{While the SM predictions provided by ATLAS are not full NNLO predictions (and do not include toponium), also the one with a rescaling of the top $p_T$ shows significant tensions with data. Furthermore, the NNLO calculation of Ref.~\cite{Mazzitelli:2020jio} displays disagreement with data at low masses and angles with older ATLAS data. Finally, Ref.~\cite{Fuks:2025toq} showed that the impact of toponium on $\Delta \phi_{e\mu}$ is limited.} 

\begin{figure*}[htb!]
\centering
\includegraphics[width=0.7\textwidth]{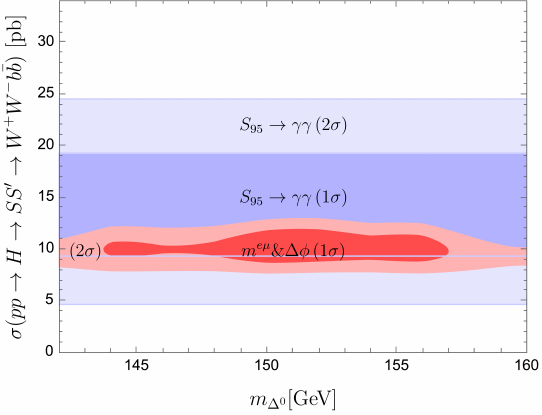}
\caption{Preferred regions from the $t\bar t$ differential distributions (red) as a function of $m_S$ and the total cross section $pp\to H\to SS^\prime\to WWbb$ assuming $S^\prime$ to be SM-like and Br$[S\to WW]=100\%$. The blue region is preferred by the $95\,$GeV $\gamma\gamma$ signal strength. Note that if one aims at explaining the $t\bar t$ differential distributions in this setup, one predicts the correct di-photon signal strength for the 95\,GeV Higgs candidate. Furthermore, the mass of $S$ determined from the leptonic $t\bar t$ distributions of $\approx 151\pm 6$\,GeV is consistent with the di-photon excesses in associated production.}
\label{ttbarHSS'}
\end{figure*}

As shown in Ref.~\cite{Banik:2023vxa}, an NP contribution of the form $pp\to H\to (S(152)\to WW)(S^\prime(95)\to b\bar b)$ improves the description of data by at least $5\sigma$ w.r.t.~to the SM for a preferred cross section of around 5pb and $m_S=151\pm7$\,GeV for $m_H\approx (250-300)$\,GeV. Furthermore, assuming that $S^\prime(95)$ has SM-like branching ratios, the preferred di-photon signal strength of the 95\,GeV excess is recovered (see Fig.~\ref{ttbarHSS'}). 

Because no excess in $ZZ\to 4\ell$ distributions at around 152\,GeV is observed, this suggests that $S$ could be the neutral component of the triplet. Furthermore, $S$ could be (mostly) a singlet, thus having SM-like branching ratios decaying dominantly to $b\bar b$, while $H$ is likely the neutral component of an $SU(2)_L$ doublet such that it can have unsuppressed dim-4 couplings to top-quarks to achieve the preferred $O(5\rm{pb})$ production cross section. Thus, a prime candidate for a UV completion explaining the 95\,GeV and 152\,GeV di-photon excesses as well as the differential $t\bar t$ (and $ttZ$) differential distributions, is the $\Delta$2HDMS, obtained by adding a singlet, a doublet and a triplet to the SM particle context~\cite{Coloretti:2023yyq}.

\section{Conclusions and Outlook}
Extensions of the Higgs sector are among the best-motivated avenues for physics beyond the SM. In these proceedings, we reviewed current indications for additional Higgs bosons at the electroweak scale, focusing on the excesses at 95\,GeV and 152\,GeV, and discussed their interpretation in extensions of the scalar sector.

The 95\,GeV candidate is supported by hints in several channels, most prominently in $\gamma\gamma$, together with weaker-than-expected limits in other final states. The 152\,GeV candidate exhibits a more distinctive pattern, pointing towards associated production via Drell-Yan and thus suggesting a non-trivial $SU(2)_L$ representation of the new Higgs. In particular, a triplet with $Y=0$ can explain the di-photon excesses ($\approx 4\sigma$). 

An interesting consequence in this model is the decay $t\to H^\pm b$ with $H^\pm\to WZ$, which contributes to $t\bar t Z$-like final states where current measurements see a slight preference for such a signal. Furthermore, the 95\,GeV and 152\,GeV excesses could be connected through processes that mimic $t\bar t$ production, potentially improving the description of differential $t\bar t$ distributions. In this context, the $\Delta$SM provides a simple and predictive framework providing a combined explanation of the 95\,GeV and 152\,GeV di-photon excesses and the potential to explain the matter-antimatter asymmetry of the Universe~\cite{Inoue:2015pza}. These excesses can be substantiated with LHC run-3 data, providing the exciting and realistic possibility of the first discovery of a  beyond-the-SM particle within this decade.

\end{document}